\documentclass[12pt,a4paper,reprint,nofootinbib,superscriptaddress,onecolumn]{revtex4-2}
\usepackage[utf8]{inputenc}
\usepackage[T1]{fontenc}
\usepackage{amsmath}
\usepackage{amssymb}
\usepackage{graphicx}
\usepackage{physics}
\usepackage{hyperref}
\hypersetup{colorlinks=true, citecolor=blue, urlcolor=blue, linkcolor=blue}

\begin{document}
\title{Electric-field driven nonequilibrium phase transitions in AdS/CFT}
\author{Daisuke Endo}
\affiliation{Department of Physics, Chuo University, Tokyo 112-8551, Japan}
\author{Yuichi Fukazawa}
\affiliation{Department of Physics, Chuo University, Tokyo 112-8551, Japan}
\author{Masataka Matsumoto}
\affiliation{Wilczek Quantum Center, School of Physics and Astronomy, Shanghai Jiao Tong University, Shanghai 200240, China}
\author{Shin Nakamura}
\affiliation{Department of Physics, Chuo University, Tokyo 112-8551, Japan}
\begin{abstract}
	We study phase transitions and critical phenomena in nonequilibrium steady states controlled by an electric field. 
	We employ the D3/D7 model in the presence of a charge density and electric field at finite temperatures.
	The system undergoes the first-order and the second-order phase transitions under the variation of the electric field in the presence of dissipation. 
	We numerically find that the critical exponents which we define for the nonequilibrium phase transition in this model take the mean-field values.
\end{abstract}
\maketitle
\tableofcontents
\section{Introduction} \label{introduction}
Extension of thermodynamics and statistical mechanics 
into nonequilibrium regime is not at all straightforward, even into a nonequilibrium steady state (NESS) which is a natural and simple extension of equilibrium states\footnote{An extension of thermodynamics into NESSs has been discussed, for example, in \cite{Oono1998steady,Sasa2006steady}. An recent proposal along this direction
with a constant heat current is found in \cite{Sasa-Nakagawa,Nakagawa2019global}.}.
In a NESS, a net current and the force acting on it, such as an electric current and electric field in a conducting system, are new macroscopic quantities compared to an equilibrium state.
Therefore, it is natural to expect that they play a crucial role in the description of the macroscopic physics of nonequilibrium systems. 
%
In general, phase transitions are important to investigate the macroscopic quantities of the systems.
In particular, critical phenomena associated with a second-order phase transition are of great interest since their universal behavior is common for a large class of systems without depending on the microscopic details.
Therefore, we believe that investigating the dependence of nonequilibrium phase transitions on current and the force acting on it would be very useful in revealing the detail-independent macroscopic nature of NESSs.


With the above motivation, we study the nonequilibrium phase transition driven by the electric field and its critical phenomena in this paper.
The corresponding phase transition driven by the electric field, for example, has been experimentally observed in the organic conductor, $\theta$-(BEDT-TTF)$_2$CsCo(SCN)$_4$ \cite{Sawano2005organic}.
The authors of \cite{Sawano2005organic} observed that the conductivity of the system discontinuously changes when the electric field acting on the system continuously varies (see, Fig.~1a in \cite{Sawano2005organic}). However, as far as the authors know, any detailed investigation of this ``nonequilibrium phase transition'' has not been done yet.
In particular, the observed transition is discontinuous in conductivity which we categorize as a first-order phase transition. A natural question is whether any second-order phase transition can be observed, and if yes, what is the universality class?

To address this issue, we employ the D3/D7 model \cite{Karch:2002sh} that is studied in the framework of the AdS/CFT correspondence (or holography).
The AdS/CFT correspondence is the equivalence between a classical gravity theory and a strongly coupled quantum gauge theory \cite{Maldacena:1998,Gubser:1998,Witten:1998}, and enables us to study nonequilibrium physics (see, for example, \cite{Hubeny:2010ry,Kundu:2019ull} and the references therein).
The dual description of the D3/D7 model is the ${\cal{N}}=4$ supersymmetric Yang-Mills theory with ${\cal{N}}=2$ hypermultiplets.
The NESS with a constant electric current in the dual field theory can be realized by considering the corresponding solutions in the bulk theory in the presence of an external electric field \cite{Karch:2007pd}.

The reason to employ the D3/D7 model is as follows. This model has the same dimensionality (namely, three spatial dimensions) as the systems investigated in \cite{Sawano2005organic}. Furthermore, the model shows negative differential conductivity \cite{Nakamura:2010zd} that is qualitatively similar to the $J$-$E$ characteristics observed in \cite{Sawano2005organic} (see, the inset of Fig. 1b in \cite{Sawano2005organic}).

The purpose of this paper is to study the nonequilibrium phase transition driven by the electric field in the D3/D7 model.
A current-driven nonequilibrium second-order phase transition has been found in this model \cite{Nakamura:2012ae} and has been investigated in \cite{Nakamura:2012ae,Matsumoto:2018ukk}. However, electric-field-driven nonequilibrium phase transitions have not yet been studied in detail in this model.
As we shall show later, we find that the system undergoes a second-order phase transition at finite charge density at a finite temperature\footnote{Here, the temperature means the temperature of the heat bath attached to the NESS.} at a finite electric field.
We investigate the critical phenomena at the critical point and determine the values of the critical exponents.
In this paper, we focus on the static critical exponents ($\beta,\delta,\gamma$) defined later because they can be defined straightforwardly in analogy with equilibrium phase transitions.
We define the order parameter by using the chiral condensate. However, we also exhibit critical phenomena for the case we regard the conductivity as the order parameter since the conductivity is easier than the chiral condensate to observe experimentally.
We numerically determine the critical exponents for each definition of the order parameter.

The rest of the paper is organized as follows.
In section \ref{setup}, we present our setup of the D3/D7 model.
In section \ref{JE}, we show the nonlinear $J$-$E$ characteristics where the electric-field driven first-order phase transition, the electric-field driven second-order phase transition, and crossover appear.
In section \ref{phase diagram}, we show the phase diagram of the nonequilibrium phase transition.
We determine the transition points of the first-order phase transitions by considering a natural extrapolation of the equilibrium free energy to the NESS. 
In section \ref{critical phenomena}, we numerically determine the values of the critical exponents.
We devote section \ref{conclusion} to conclusion and discussions. We make a comment on our computation of the current density in Appendix A.

\section{Setup} \label{setup}
In this section, we briefly present the D3/D7 model at finite temperature in the presence of the charge density and electric field \cite{Karch:2007pd}.
We consider a (3+1)-dimensional SU($N_{c}$) ${\cal{N}}=4$ supersymmetric Yang-Mills (SYM) theory with ${\cal{N}}=2$ hypermultiplet (HM) in the large-$N_{c}$ limit with a large 't~Hooft coupling $\lambda=g_{\rm YM}^{2}N_{c}$. 
At finite temperatures, the degrees of freedom of the SYM sector are $O(N_{c}^2)$ whereas that of the HM sector is $O(N_{c})$: the SYM sector plays the role of heat bath against the possible perturbations of the HM sector by virtue of the large-$N_{c}$ limit. 

In this paper, we apply a constant (and homogeneous) external electric field acting on the $U(1)$ global charge in the HM sector. Since the charges are interacting with the heat bath made of the SYM sector, we can realize a NESS where we have a constant flow of the current along the constant external electric field. The particles in the HM sector have a finite mass that plays the role of the mass gap of the charged particles. Therefore, the system can be either an insulator or a conductor depending on the magnitude of the mass gap. Even with a finite value of the mass gap, the system exhibits a finite conductivity for a large enough electric field owing to the pair creation of the positive and the negative charges. In this sense, the system is understood as a field-theory model of strongly correlated insulators. In our analyses, we introduce a chemical potential of the charge in order to realize a finite charge density.


This theory is conjectured to be dual to the D3/D7 model \cite{Karch:2002sh}, whose geometry at finite temperature is the five-dimensional AdS-Schwarzschild black hole times $S^{5}$. 
The metric of the geometry is explicitly given by
\begin{equation}
	\dd s^{2} = \frac{L^{2}}{u^{2}}\left( -f(u)\dd t^{2} + \dd\vec{x}^{2} +\frac{\dd u^{2}}{f(u)} \right) + L^{2}\dd\Omega_{5}^{2}, 
\end{equation}
with
\begin{equation}
		\dd\Omega_{5}^{2} = \dd\theta^{2} + \sin^{2}\theta \dd \psi^{2} + \cos^{2}\theta \dd \Omega_{3}^{2},
\end{equation}
where $f(u)=1-u^{4}/u_{\rm H}^4$.
$(t,\vec{x})$ are (3+1)-dimensional spacetime coordinates in the field theory and $u$ is the radial coordinate.
The boundary and the black hole horizon are located at $u=0$ and $u=u_{\rm H}$, respectively.
The Hawking temperature $T= 1/\pi u_{\rm H}$ gives the heat bath temperature.

The dynamics of the D7-brane is given by the Dirac-Born-Infeld (DBI) action:
\begin{eqnarray}
	S_{D7} &=& -T_{D7} \int \dd^{8}\xi \sqrt{-\det \left( \tilde{g}_{ab} + 2\pi l_{\rm s}^{2} F_{ab} \right)} \nonumber \\ 
	&=& -T_{D7}L^{8} \int \dd^{8}\xi \sqrt{-\det \left( g_{ab} + 2\pi l_{\rm s}^{2} L^{-2} F_{ab} \right)},
\end{eqnarray}
where $T_{D7}$ is the tension of the D7-brane given by $T_{D7}=(2\pi)^{-7}l_{\rm s}^{-8} g_{\rm s}^{-1}$ with the string length $l_{\rm s}$ and the string coupling $g_{\rm s}$.
$\tilde{g}_{ab}=L^{2}g_{ab}$ is the induced metric of the D7-brane, and $F_{ab}\equiv \partial_{a}A_{b} - \partial_{b}A_{a}$ is the field strength of the $U(1)$ gauge field on the D7-brane.
The indices $a$ and $b$ label the worldvolume coordinates $\xi^{a}$.
In the following, we set $2\pi l_{s}^{2} L^{-2}=1$ so that $\lambda=(2\pi)^{2}/2$.

For our purposes, we employ the following ansatz of the fields,
\begin{equation}
	\theta= \theta(u), \hspace{1em} A_{t} = A_{t}(u), \hspace{1em} A_{x} = -Et + h(u),
\end{equation}
where $E$ corresponds to the external electric field. 
Then, the induced metric is explicitly given by
\begin{equation}
	\dd s^{2}_{D7} = \frac{1}{u^{2}} \left( -f(u) \dd t^{2} + \dd \vec{x}^{2} \right) + \left(\frac{1}{u^{2}f(u)}+ \theta'(u)^{2} \right) \dd u^{2}+\dd \Omega_{3}^{2},
\end{equation}
where the prime denotes the derivative with respect to $u$.
Thus, the DBI action becomes
\begin{equation}
	S_{D7} = - {\cal{N}} V_{4} \int \dd u \cos^{3}\theta g_{xx} \sqrt{-g_{tt}g_{xx}g_{uu} - \left(g_{xx}A_{t}'^{2}+g_{uu}E^{2} +g_{tt}h'^{2} \right)},
\end{equation}
where $V_{4} = \int \dd t\dd x\dd y\dd z$ and 
\begin{equation}
	{\cal{N}} = T_{D7}(2\pi^{2})L^{8} =\frac{N_{c}}{\lambda}.
\end{equation}
Here, the 't~Hooft coupling is given by $\lambda=2\pi g_{s}N_{c}$
in our convention.\footnote{We employ a convention of $g_{\rm YM}^{2}=2\pi g_{s}$ and $\lambda=g_{\rm YM}^{2}N_{c}$.}
Now, we set ${\cal{N}}=1$, for simplicity, and $N_{c}=\lambda=(2\pi)^{2}/2$ in the rest of this paper.

Since the action contains the only derivative terms of the gauge fields, the following quantities are conserved along $u$ direction,
\begin{align}
	\rho &\equiv \frac{\delta S_{D7} }{\delta A_{t}'} = \frac{-g_{xx}^{2}A_{t}'\cos^{3}\theta}{\sqrt{\xi g_{uu} -g_{tt}h'^{2}-g_{xx}A_{t}'^{2} }},\\
	J &\equiv \frac{\delta S_{D7} }{\delta A_{x}'} = \frac{-g_{tt}g_{xx}h'\cos^{3}\theta}{\sqrt{\xi g_{uu} -g_{tt}h'^{2}-g_{xx}A_{t}'^{2} }},
\end{align}
where $\xi= -g_{tt}g_{xx}-E^{2}$.
Here, $\rho$ and $J$ are the charge density and the current density along the $x$ direction, respectively.
There is a location $u=u_{*}$ in the range of $0 \leq u \leq u_{\rm H}$ that satisfies $\xi=0$ because $-g_{tt}$ monotonically decreases toward the horizon.
$\rho$ and $J$ are given by functions of $A_{t}'$ and $h'$.
However, one finds that $g_{xx}J^{2}+g_{tt}\rho^{2}$ is independent of $A_{t}'$ and $h'$ at $u=u_{*}$. Hence, we obtain
\begin{equation}
	J^{2} = -\frac{g_{tt}}{g_{xx}}\rho^{2} - g_{tt} g_{xx}^{2} \cos^{6}\theta(u_{*}),
 \label{eq:current}
\end{equation}
where all the metric components are evaluated at $u=u_{*}$.\footnote{Note that our derivation of $J$ for finite $\rho$ here is an extension of the method in \cite{Hashimoto:2014yza,Ishigaki:2021vyv} for $\rho=0$. The details are given in Appendix \ref{AppendixA}.}
The location $u_{*}$ is referred to as the effective horizon because it forms the causal boundary for the dynamics on the D7-brane \cite{Kim:2011qh,Seiberg:1999vs}. 

From the equations of motion, each field can be expanded near the boundary as
\begin{align}
	\theta(u) &= m u + \theta_{2} u^{3} + \cdots, \label{eq:thetaasym}\\
	A_{t}(u) &= \mu -  \frac{\rho}{2}u^{2} +\cdots, \\
	h(u) &= b +  \frac{J}{2}u^{2} +\cdots,
\end{align}
where $m$ and $\mu$ correspond to the mass of the charged particles in the HM sector and the chemical potential, respectively.
We assume that the source term $b$ for $h(u)$ vanishes.
The normalizable mode of $\theta(u)$ is related to the chiral condensate via
\begin{equation}
	\left<\bar{q}q \right> = - \left( -2 \theta_{2} + \frac{m^{3}}{3} \right).
\end{equation}
The normalizable modes of $A_{t}(u)$ and $h(u)$ are corresponding to the conserved quantities $\rho$ and $J$ mentioned above.

For convenience, we perform the Legendre transformation with respect to the gauge fields $A_{t}$ and $A_{x}$.
The transformed DBI action is given by
\begin{align}
	\tilde{S}_{D7} &= S_{D7} - \int \dd u \left(\frac{\delta S_{D7}}{\delta A'_{t}}A'_{t}+ \frac{\delta S_{D7}}{\delta A'_{x}}A'_{x} \right) = - V_{4} \int \dd u \tilde{{\cal{L}}}_{D7},
\end{align}
where
\begin{align}
    \tilde{{\cal{L}}}_{D7} = \cos^{3}\theta \sqrt{\frac{-g_{uu}}{g_{tt} g_{xx}}}\sqrt{\left(g_{tt}g_{xx} +E^{2} \right) \left(  g_{tt}g_{xx}^{3}\cos^{6}\theta + g_{tt}\rho^{2} + g_{xx} J^{2} \right)}.
\end{align}
Since the action is written by the variables ($\theta,\theta', \rho, J$), 
we can obtain the equation of motion for $\theta(u)$ with $\rho$ and $J$ fixed from the transformed action.
We will not present the explicit form of the equation of motion because it is cumbersome and not very illuminating.

\section{Nonlinear characteristics} \label{JE}
In this section, we study the $J$-$E$ characteristics obtained from the solutions to the equation of motion.
There are three different types of solutions depending on the configuration of the D7-brane.
The solutions that do not reach the black hole horizon are referred to as the {\it Minkowski embeddings}, whereas the solutions that fall into the black hole are called the {\it black hole embeddings}.
The D7-brane in the black hole embedding goes across the effective horizon in the presence of the electric field.
The other type of solution is called the {\it critical embedding}.
The critical embedding is between the Minkowski embedding and the black hole embedding, and forms a conical singularity at the effective horizon.
In the dual field theory, the Minkowski embedding and the black hole embedding are in the insulating phase and the conducting phase, respectively.
Since we are interested in the nonequilibrium steady state with a constant current, we focus only on the black hole embeddings.

For this purpose, we numerically solve the equation of motion for $\theta(u)$ with given ($T,\rho,E$) under the following boundary conditions specified at the effective horizon.
First, we specify the value of $\theta(u_{*})$, which corresponds to choosing the value of $J$.
The other boundary condition that constrains the value of $\theta^{\prime}(u_{*})$ is obtained from the equation of motion at the effective horizon.
Here, we employ the shooting method for the numerical calculation.
After we obtain the numerical solution, we can read off the mass of the charge carriers from the asymptotic form of $\theta(u)$ near the AdS boundary from (\ref{eq:thetaasym}).
The system is invariant under the scale transformation,
\begin{align}
	&(t,\vec{x},u) \to (a t, a \vec{x},a u), \hspace{0.5em} m\to m/a, \hspace{0.5em} \theta_{2} \to \theta_{2} /a^{3}, \\
	&E \to E/a^{2}, \hspace{0.5em} J \to J/a^{3}, \hspace{0.5em} \mu \to \mu/a, \hspace{0.5em} \rho \to \rho/a^{3},
\end{align}
where $a$ is an arbitrary constant.
Using this scale invariance, the scale-free parameters which characterize the system are given by $(T/m, \rho/m^{3}, E/m^{2})$.

\begin{figure}[tbp]
	\centering
	\includegraphics[width=10cm]{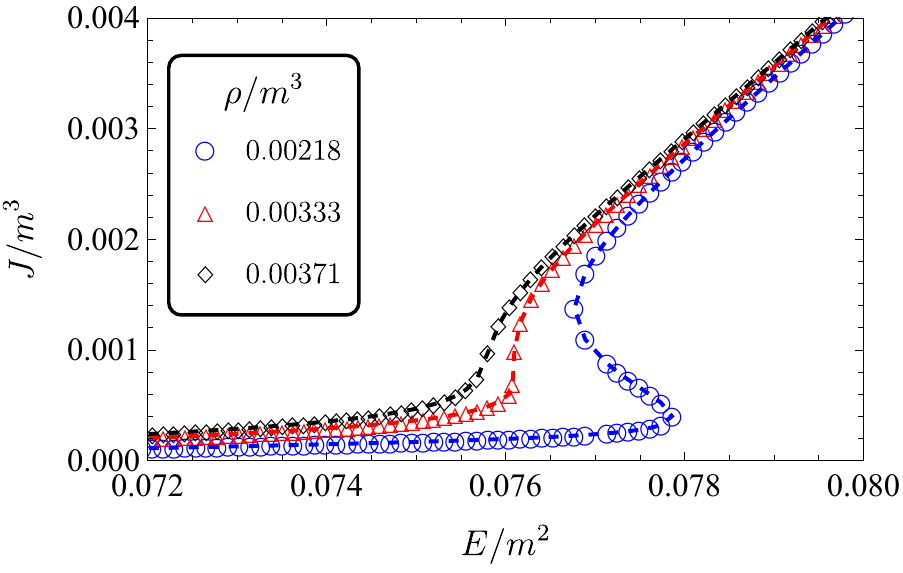}
	\caption{$J$-$E$ characteristics for several $\rho/m^3$ with $T$ fixed.}
	\label{fig:JE}
\end{figure}

If we specify the parameters $(T/m, \rho/m^{3},E/m^{2})$, we obtain the value of $J/m^{3}$ by solving the equations of motion.
Collecting the solutions for various values of $E/m^{2}$ with the other parameters fixed, we obtain the $J$-$E$ characteristics for the given value of $(T/m, \rho/m)$.
Figure \ref{fig:JE} shows typical plots of the $J$-$E$ characteristics for several values of $\rho/m^{3}$ with $\pi T/m=1.073$.
As indicated there, $J/m^{3}$ is a multi-valued function of $E/m^{2}$ for $\rho/m^{3}=0.00218$. 
Since only one of the multiple values of $J/m^{3}$ should be realized at a given $E/m^{2}$ in the physical systems, the value of $J/m^{3}$ must jump somewhere in the multi-valued region when we vary $E/m^{2}$.
We regard this discontinuous jump as the first-order phase transition of the systems driven by the electric field.
The method of determining the transition point is described in the next section.
We also find that $J/m^{3}$ becomes a single-valued function when $\rho/m^{3}$ is large enough along with $T/m$ fixed: the first-order phase transition changes into the second-order phase transition and then into the crossover if we increase $\rho/m^{3}$ with $T/m$ fixed.

The emergence of the critical point, the point where the foregoing second-order phase transition occurs, 
can be understood as follows.
The charge density created by the presence of a chemical potential contributes to the positive differential conductivity, characterized by $\frac{\partial J}{\partial E}>0$, that gives a monotonically increasing current when we increase the electric field. 
On the other hand, the current is also carried by the particles that are produced through the pair creation of the positive and the negative charge carriers induced by the electric field. It has been discussed in \cite{Ishigaki:2020coe} that they provide the negative differential conductivity,\footnote{The negative differential conductivity has been observed in the D3/D7 model in \cite{Nakamura:2010zd} at zero charge density.} characterized by $\frac{\partial J}{\partial E}<0$, at small current density.
As a result of the competition of these two contributions, the critical point where  $\frac{\partial E}{\partial J}=0$ (hence $\frac{\partial J}{\partial E}$ is divergent) emerges at the critical values of ($T/m, \rho/m^{3}$).

\section{Phase Diagram}	\label{phase diagram}
In holography, the free energy of an equilibrium system is given as the Hamiltonian in the gravity dual.
This idea has been extended to the systems of NESS in the previous studies \cite{Nakamura:2012ae,Matsumoto:2018ukk,Imaizumi:2019byu,Matsumoto:2022psr}.
In these studies, the authors considered the $\rho=0$ case and employed a Legendre-transformed Hamiltonian in such a way that $J$ is a control parameter.
To be concrete, 
the corresponding Hamiltonian (per unit volume) is given by
\begin{equation}
	\tilde{H} = \int_{0}^{u_*} du \tilde{{\cal{H}}} =\int_{0}^{u_*}  du \left[ E\frac{\partial \Bar{{\cal{L}}}_{D7}}{\partial E}  - \Bar{{\cal{L}}}_{D7}\right],
 \label{eq:tildeH}
\end{equation}
where
\begin{equation}
\Bar{{\cal{L}}}_{D7} = {\cal{L}}_{D7}+\frac{\partial {\cal{L}}_{D7}}{\partial A_{x}^{\prime}}A_{x}^{\prime}
\end{equation}
is the Legendre-transformed Lagrangian density that is a function of $(\partial_{t}{A}_{x}=E, J, \theta,\theta')$. The integral is evaluated from the boundary to the effective horizon. In this case, $\tilde{H}$ is not a function of $E$, but of $J$.
This Hamiltonian has been used to define the location of the first-order phase transitions on the phase diagram in NESS.
The definition of the first-order phase transition line affects the obtained value of the critical exponent $\beta$ since it is defined as how the order parameter scales when the control parameter approaches the critical point along the first-order phase transition line. 
The critical exponents, including $\beta$, obtained in the previous works showed a consistent behavior exhibiting the mean-field theory values.\footnote{See also a more precise statement in section \ref{conclusion}.} 

Let us extend the above idea to the present system and see how it works.
In this study, we use $E$, but not $J$, as a control parameter because we are dealing with the electric-field driven phase transition. We also introduce the charge density $\rho$ as a control parameter.
Thus, we employ the following quantity as the free energy of the NESS.
\begin{equation}
	F = -\int_{0}^{u_{*}} du \left({\cal{L}}_{D7}+\frac{\partial {\cal{L}}_{D7}}{\partial A'_{t}}A'_{t}\right),
\end{equation}
where the integral is evaluated from the boundary $u=0$ to the effective horizon $u=u_{*}$. Note that we removed the Legendre transformation that makes $J$ a control parameter from (\ref{eq:tildeH}), and performed the Legendre transformation that makes $\rho=-\frac{\partial {\cal{L}}_{D7}}{\partial A'_{t}}$ a control parameter. 
Here, the Lagrangian must be renormalized because the integral with respect to $u$ is divergent at the boundary.
According to the holographic renormalization in the probe brane model \cite{Karch:2005ms,Karch:2007pd}, we introduce the following counterterms,
\begin{equation}
 	L_{\rm count} \equiv L_{1}+L_{2} + L_{f} +L_{F},
\end{equation}
where $\varepsilon$ is the cutoff near the boundary and each counterterm is given by
\begin{align}
	&L_{1} = \frac{1}{4} \sqrt{-\gamma}, \hspace{1em} L_{2} = -\frac{1}{2} \sqrt{-\gamma} \theta(\varepsilon)^{2}, \\
	&L_{f} = \frac{5}{12}\sqrt{-\gamma} \theta(\varepsilon)^{4}, \hspace{1em} L_{F} = \frac{1}{2}\sqrt{-\gamma} E^{2} \log \Lambda\varepsilon,
\end{align}
where $\gamma=\det \gamma_{ij}$ and $\gamma_{ij}$ is the induced metric at the boundary $u=\varepsilon$.
The factor $\Lambda$ is introduced so that $\Lambda \varepsilon$ becomes dimensionless and gives a finite contribution to $F$ depending on the choice of $\Lambda$. However, the ambiguity of $\Lambda$ does not 
contribute to our analysis since $L_{F}$ drops from the difference of $F$ at a common $E$.
Adding the counterterm to the Lagrangian, we obtain the renormalized free energy for the NESS,
\begin{equation}
	F_{\rm ren} = -\left[ \int^{u_{*}}_{\varepsilon} du \left({\cal{L}}_{D7}+\frac{\partial {\cal{L}}_{D7}}{\partial A'_{t}}A'_{t}\right)+ L_{\rm count} \right].
\end{equation}

\begin{figure}
	\centering
	\includegraphics[width=10cm]{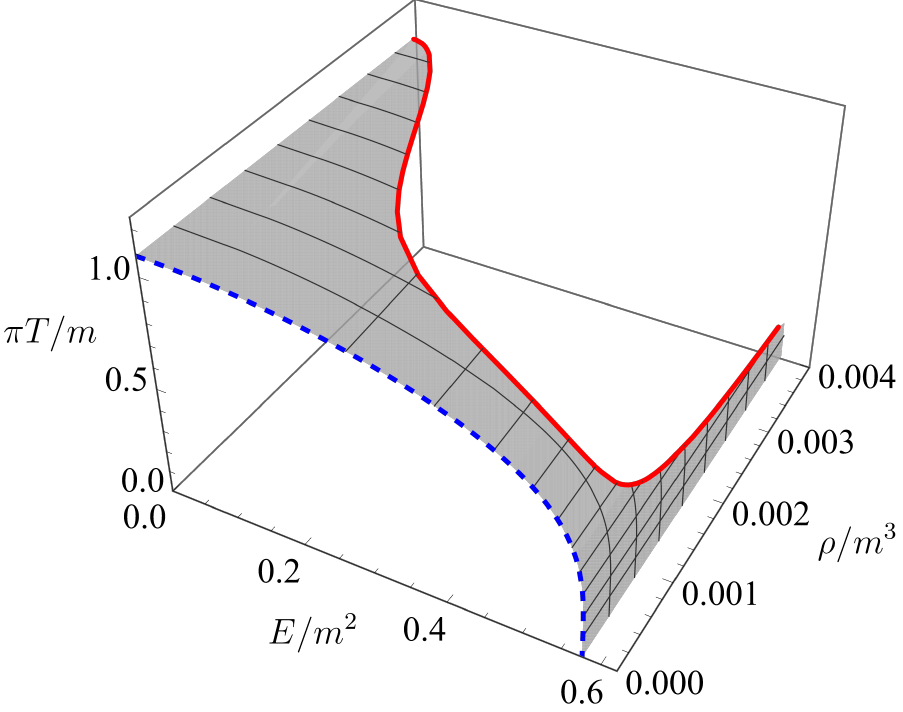}
	\caption{The three-dimensional phase diagram for the nonequilibrium phase transition. The red solid curve denotes the second-order phase transition points. The blue dashed curve denotes the first-order phase transition points at $\rho=0$. The gray surface between them denotes the first-order phase transition points.}
	\label{fig:phase}
\end{figure}

Evaluating the free energy for each parameter $(T,\rho,E)$, we obtain the phase diagram for our nonequilibrium phase transition as shown in figure \ref{fig:phase}.
The red solid curve denotes the second-order phase transition points.
We refer to this curve as the critical line in this paper.
The blue dashed curve denotes the first-order phase transition points at zero charge density, corresponding to the phase diagram shown in \cite{Ishigaki:2021vyv}.
The gray-shaded surface denotes the first-order phase transition points.
We find that the critical line ends at some point of the $(T, \rho)$
plane in the limit of $E\to 0$.
This means that the differential conductivity is divergent at zero electric field, $\left. \partial J / \partial E \right|_{E=0} \to \infty$ at this point since $\partial J / \partial E$ is divergent at the critical point. 
 (See figure \ref{fig:JE}.)
We also find that the critical line approaches the critical value of the electric field $E_{\rm crit}/m^{2} \approx 0.57$ in the limit of $T\to 0$.
Here, $E_{\rm crit}$ is the value of $E$ where the D7-brane barely touches the effective horizon.\footnote{Note that $E_{\rm crit}$ is different from $E_c$, which we will use in Section \ref{critical phenomena}, defined as the critical value of the electric field for the second-order nonequilibrium phase transitions.}
This behavior can be understood from the fact that the second-order phase transition emerges due to the competition between the positive differential conductivity and the negative differential conductivity.
Since the negative differential conductivity appears for $E \lesssim E_{\rm crit}$ \cite{Nakamura:2010zd}, we expect to find the phase transitions occur only at $E \lesssim E_{\rm crit}$.

\section{Critical Phenomena} \label{critical phenomena}
In this section, we study the critical phenomena at the critical points on the critical line.
For this purpose, we define the critical exponents $\beta$, $\delta$ and $\gamma$ as follows:
\begin{align}
	\Delta\phi \propto |\kappa-\kappa_{c}|^{\beta}, \hspace{0.5em} |\phi-\phi_{c}| \propto |E-E_{c}|^{1/\delta}, \hspace{0.5em} 
	\chi \propto |\kappa - \kappa_{c}|^{-\gamma}, 
\end{align}
where $\kappa$ is a control parameter such as $T$ or $\rho$. The choice of the control parameter $\kappa$ depends on the path along which we approach the critical point on the  three-dimensional phase diagram figure \ref{fig:phase}. 
In this paper, we consider two cases where $\kappa=T$ with $\rho$ fixed and $\kappa = \rho$ with $T$ fixed.
$\phi$ is a quantity from which we define the order parameter $\Delta\phi$, where $\Delta\phi$ is the difference of $\phi$ between the two phases when the first-order phase transition occurs. 
$\chi$ is the susceptibility defined by $\chi = \partial\phi / \partial E$. 
The subscript $c$ denotes the values of each parameter at the critical point.
Note that the critical exponent $\delta$ is defined at $\kappa=\kappa_{c}$ and $\gamma$ can be evaluated both in $\kappa>\kappa_{c}$ and $\kappa<\kappa_{c}$. 

In our study, we choose the chiral condensate $\expval{\bar{q}q}$ as $\phi$. However, we also study the critical phenomena by employing conductivity $\sigma = J/E$ for $\phi$ in place of the chiral condensate. The reason is that we can observe the conductivity in experiments easier than the chiral condensate. As we shall see, we find the critical phenomena and the critical exponents are well detected by using conductivity as well as the chiral condensate.

Note that we have chosen the electric field as the external field that plays the role of the magnetic field in the Ising model. The reason is that we want to compare our result with the phase transition observed in \cite{Sawano2005organic} where hysteresis appears in the continuous variation of the electric field. 

\begin{figure}[tbp]
	\centering
	\includegraphics[width=7cm]{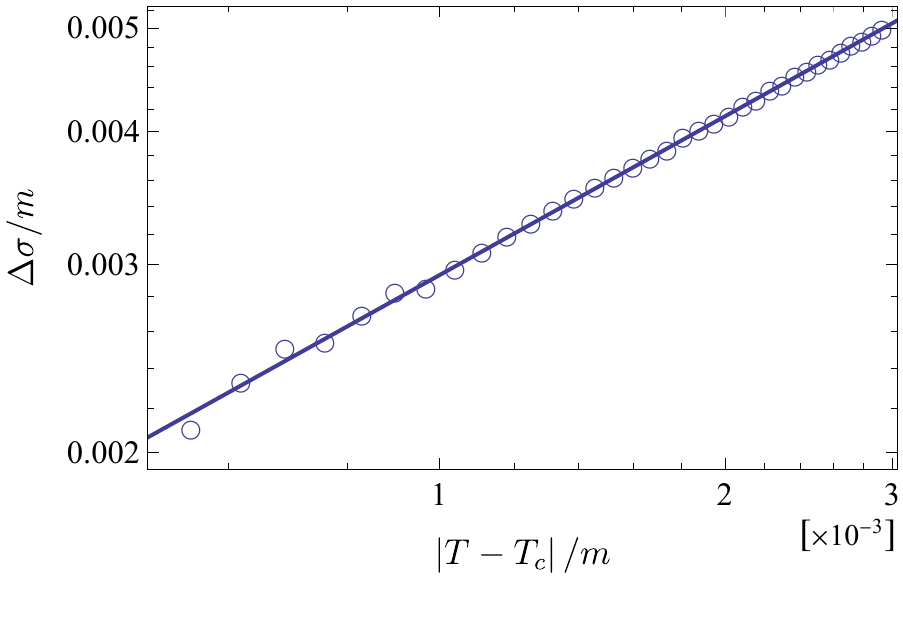}
	\includegraphics[width=7cm]{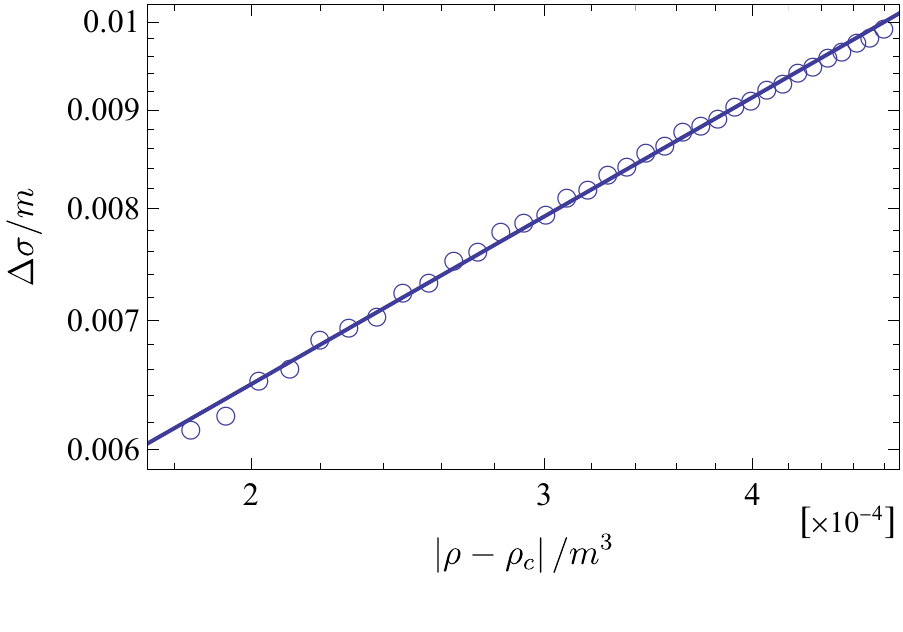}
	\includegraphics[width=7cm]{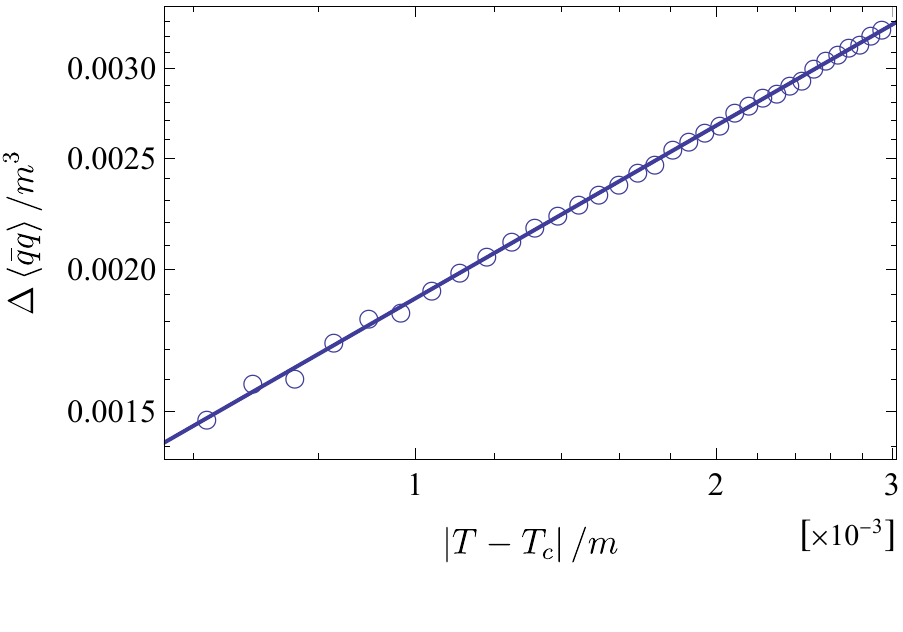}
	\includegraphics[width=7cm]{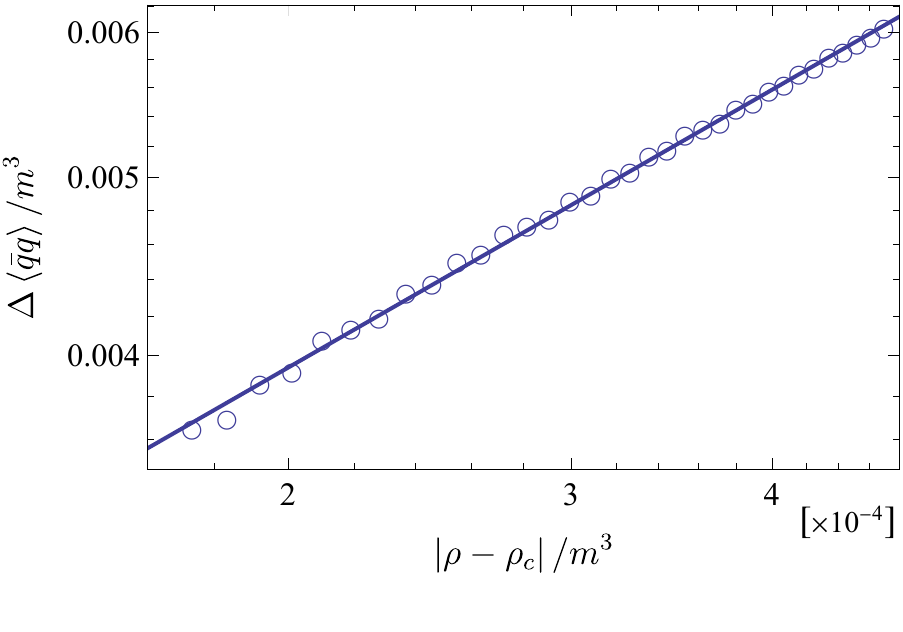}
	\caption{The critical behaviors of the order parameter.
 We show the results with the conductivity in the upper panels and those with the chiral condensate in the lower panels. The left (right) panels show the results with  $\kappa=T$ ($\kappa = \rho$). 
 The open circles and the solid lines denote the numerical plots and the fitting results, respectively. All the plots are shown in a log-log scale.}
	\label{fig:beta}
\end{figure}

In figure \ref{fig:beta}, we show the critical behaviors of $\Delta\phi$ with respect to $|\kappa-\kappa_{c}|$ for each choice of parameters. Note that all the data are normalized by using $m$. (This normalization follows in other figures as well.)
In the upper and lower panels of figure \ref{fig:beta}, we choose the conductivity and chiral condensate as the order parameter, respectively.
In the left and right panels of figure \ref{fig:beta}, we choose $\kappa = T/m$ with $\rho/m^{3}=0.00218$ fixed and $\kappa = \rho/m^{3}$ with $T/m=1.073$ fixed, respectively.
From the fittings, we find $\beta\approx0.495$ (upper left), $\beta\approx0.494$ (upper right), $\beta\approx0.503$ (lower left), and $\beta\approx0.503$ (lower right), respectively.
All of these results agree with the values of $\beta$ in the Landau theory, that is, $\beta = 1/2$.

\begin{figure}[tbp]
	\centering
	\includegraphics[width=7cm]{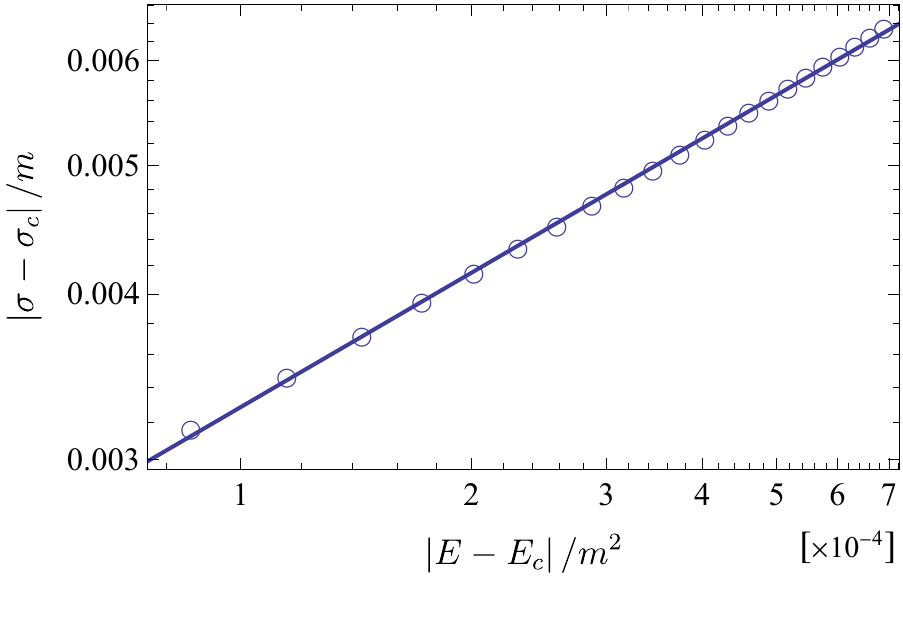}
	\includegraphics[width=7cm]{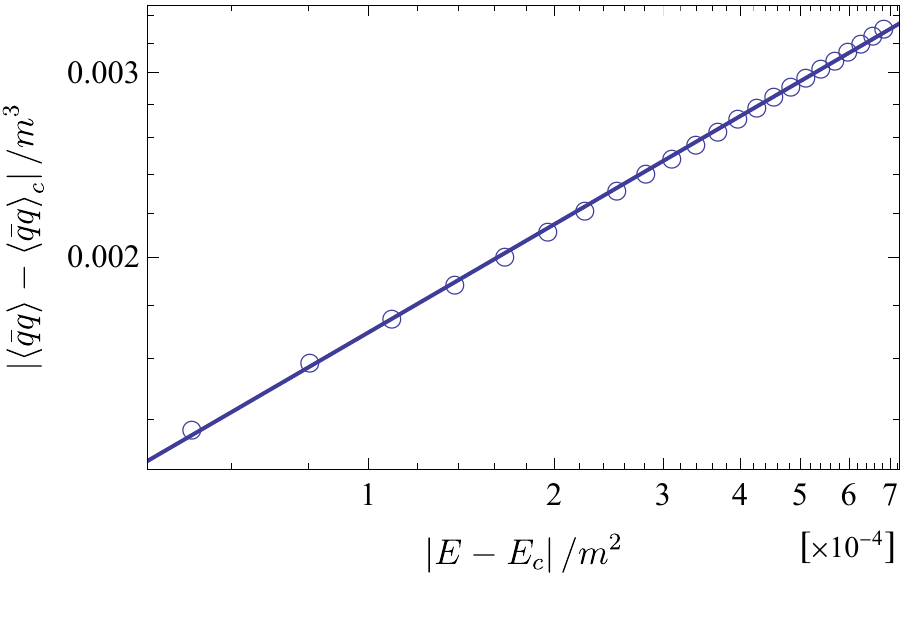}
	\caption{The critical behaviors of $|\phi-\phi_{c}|$ with respect to $|E-E_{c}|$ at the critical point. The left (right) panel shows the results with the conductivity (the chiral condensate). 
	The open circles and the solid lines denote the numerical plots and the fitting results, respectively.}
	\label{fig:delta}
\end{figure}

In figure \ref{fig:delta}, we show the critical behaviors of the order parameter with respect to the external field at the critical point. 
The left and right panels in figure \ref{fig:delta} show the critical behaviors of the conductivity and chiral condensate with respect to the electric field at the critical point, respectively.
From the fittings, we find that $1/\delta \approx 0.337$ (left panel) and $1/\delta \approx 0.343$ (right panel), respectively.
These values of $\delta$ also agree with those in the Landau theory, that is, $\delta=3$.

\begin{figure}[tbp]
	\centering
	\includegraphics[width=7cm]{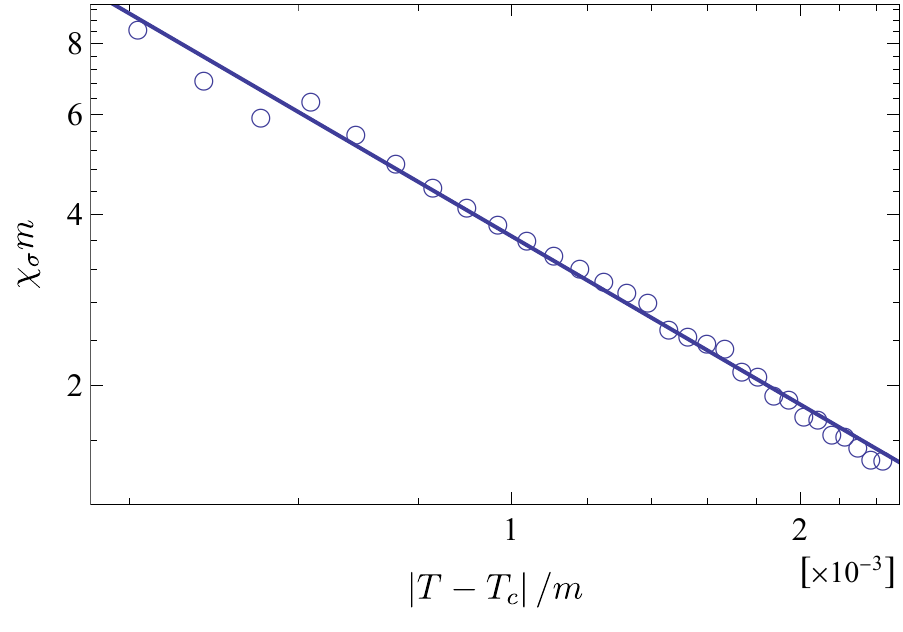}
	\includegraphics[width=7cm]{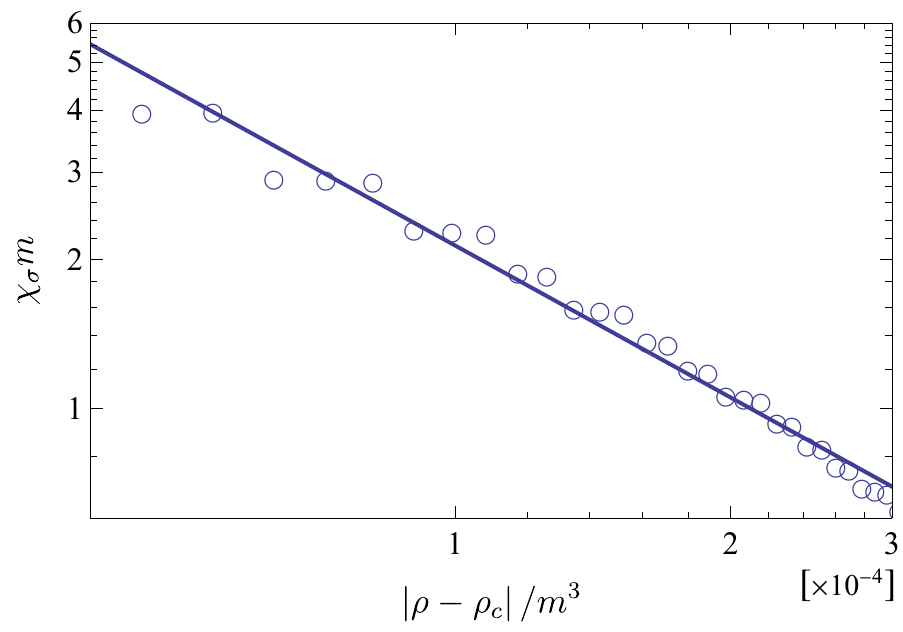}
	\includegraphics[width=7cm]{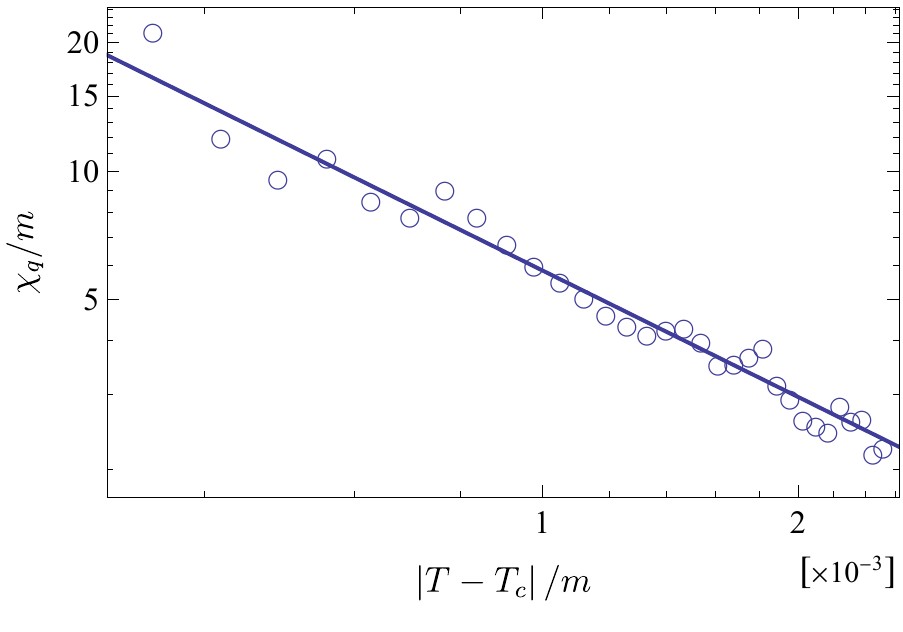}
	\includegraphics[width=7cm]{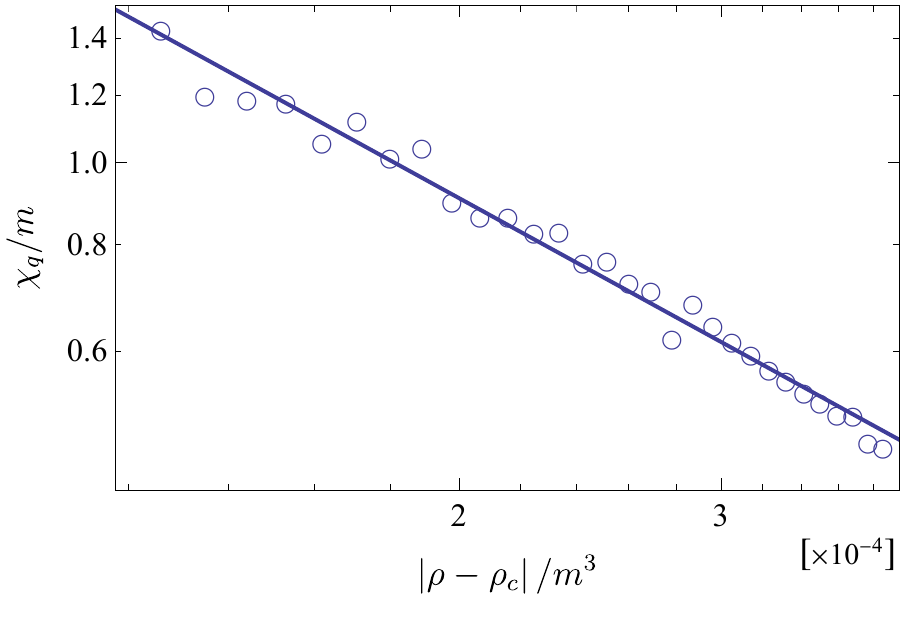}
	\caption{The critical behaviors of the susceptibilities in the first-order phase transition region. The left (right) panels show the results with the susceptibility of the conductivity (the chiral condensate). The open circles and the solid lines denote the numerical plots and the fitting results, respectively.}
	\label{fig:gamma}
\end{figure}

The critical exponent $\gamma$ can be determined by the divergent behaviors of the susceptibilities.
We define the two different susceptibilities by
\begin{equation}
	\chi_{\sigma} = \frac{\partial \sigma}{\partial E}= \frac{1}{E}\left(\frac{\partial J}{\partial E} - \frac{J}{E} \right), \hspace{0.5em} \chi_{q} = \frac{\partial \expval{\bar{q}q}}{\partial E}.
\end{equation}
We study the critical behaviors of the susceptibilities in 
$\kappa>\kappa_{c}$ and $\kappa < \kappa_{c}$, 
corresponding to the first-order phase transition and the crossover region.
In figure \ref{fig:gamma}, we show the critical behaviors of the susceptibilities in the first-order phase transitions region.
From the fitting, we find $\gamma\approx 1.020$ (upper left), $\gamma\approx0.985$ (upper right), $\gamma\approx0.960$ (lower left), and $\gamma\approx0.986$ (lower right) respectively.
We find that the values of $\gamma$ in the first-order phase transition region agree with those in the Landau theory, that is $\gamma=1$.

\begin{figure}[tbp]
	\centering
	\includegraphics[width=7cm]{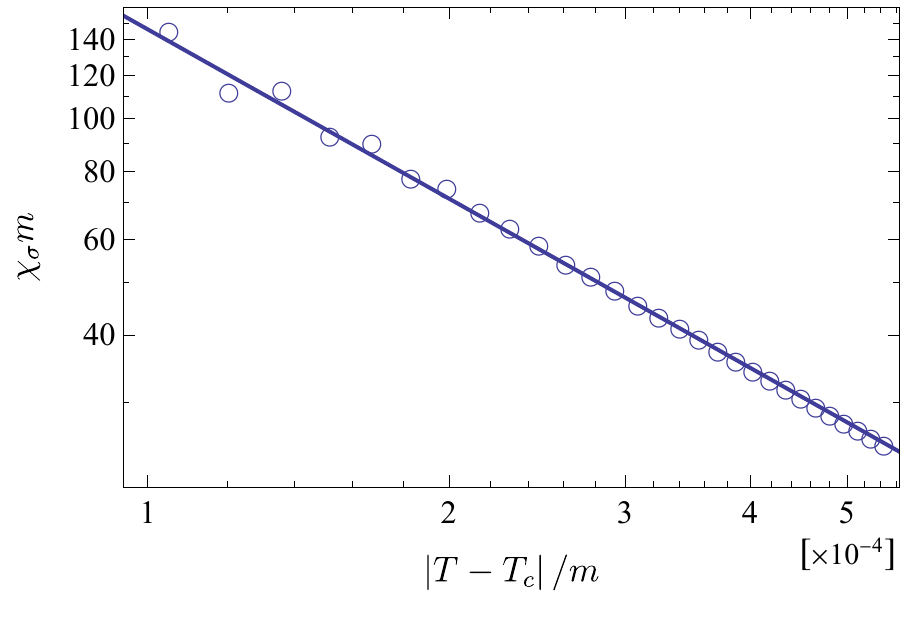}
	\includegraphics[width=7cm]{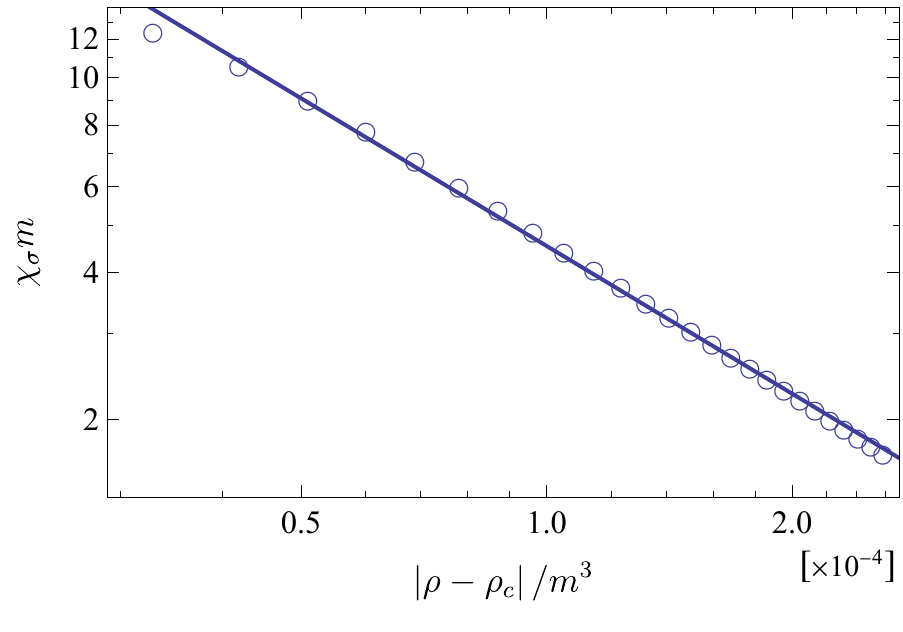}
	\includegraphics[width=7cm]{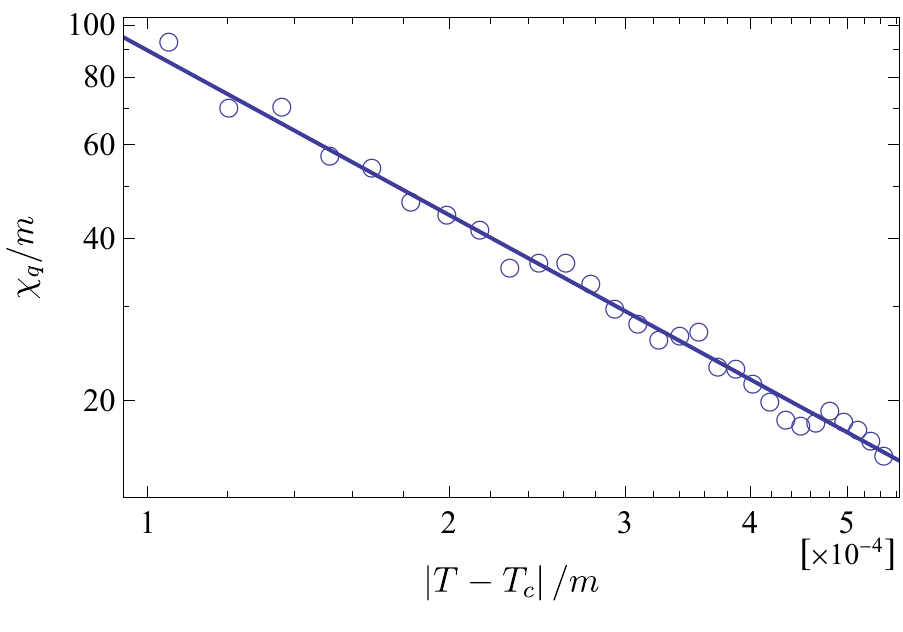}
	\includegraphics[width=7cm]{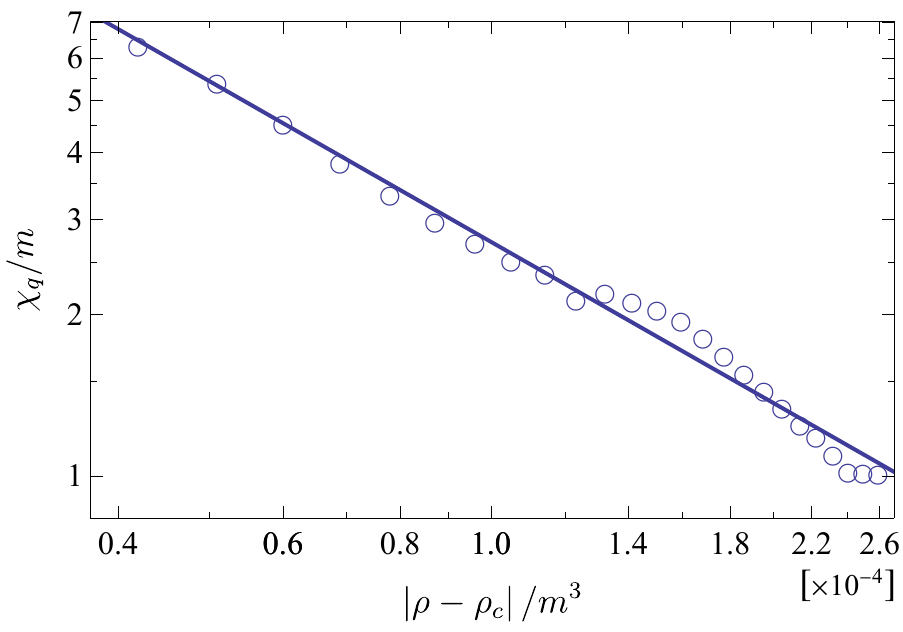}
	\caption{The critical behaviors of the susceptibilities in the crossover region. The left (right) panels show the results with the susceptibility of the conductivity (the chiral condensate). The open circles and the solid lines denote the numerical plots and the fitting results, respectively.}
	\label{fig:gammaC}
\end{figure}

In figure \ref{fig:gammaC}, we show the critical behaviors of the susceptibilities in the crossover region.
From the fitting, we find $\gamma\approx 1.003$ (upper left), $\gamma\approx1.034$ (upper right), $\gamma\approx0.996$ (lower left), and $\gamma\approx1.016$ (lower right) respectively.
We find that the values of $\gamma$ in the crossover region also agree with those in the Landau theory.

In summary, we numerically obtain the critical exponents ($\beta,\delta,\gamma$) in the electric-field driven nonequilibrium phase transitions as shown in table \ref{tab:summary}.
\begin{table}
	\caption{\label{tab:summary}Critical exponents.}
	\centering
		\begin{tabular}{p{15mm}cc|cc}
			\hline \hline
			& $T$-fixed & & \hspace{0.2em}$\rho$-fixed& \\
			 & {$\sigma$} & {$\expval{\bar{q}q}$} & {$\sigma$} & {$\expval{\bar{q}q}$}  \\ \hline
			$\beta$  & 0.495 & 0.503 & 0.494 & 0.503 \\
			$\delta$ & 2.967 & 2.915 & 2.967 & 2.915  \\
			$\gamma_{\rm 1st}$  & 1.020 & 0.960 & 0.985 & 0.986 \\
			$\gamma_{\rm cross}$  & 1.003 & 0.996 & 1.034 & 1.016 \\
			\hline \hline
		\end{tabular}
\end{table}

\section{Conclusion and Discussions} \label{conclusion}
In this paper, we studied the nonequilibrium phase transition driven by the electric field in the framework of the AdS/CFT correspondence.
We employed the D3/D7 model at finite charge densities and finite temperatures in the presence of the external electric field.
Our results show that the system undergoes the nonequilibrium phase transition due to the non-linear $J$-$E$ characteristics.
We found that there are regions of the first-order phase transition and crossover, and the critical line in between where the second-order phase transition occurs in the parameter space in the phase diagram (figure \ref{fig:phase}).
We investigated the critical phenomena at the critical points and found that the obtained values of the critical exponents ($\beta,\delta,\gamma$) agree with those in the Landau theory.

The agreement of the critical exponents with those in the Landau theory of equilibrium systems implies that our nonequilibrium phase transition can be described by an effective theory that has the same structure as the Landau theory.
The observation that the critical phenomena in the D3/D7 model agree with those of the Landau theory has been given in \cite{Matsumoto:2022psr} for thermal equilibrium and in \cite{Nakamura:2012ae,Matsumoto:2018ukk,Imaizumi:2019byu} for nonequilibrium circumstances.
This agreement may be understood as the suppression of fluctuations near the critical point owing to the large-$N_{c}$ limit.
However, deviations from the mean-field values of the critical exponents have been observed in the current-driven tricritical point \cite{Matsumoto:2022nqu}.
It would be interesting to see whether we have a tricritical point for the electric-field driven systems and study the critical phenomena there if it is the case.

Let us make a comment on a possible connection between our results and experiments.
We have shown that the present model provides a concrete theoretical example that reproduces an electric-field driven first-order phase transition similar to that has been observed in the organic conductor \cite{Sawano2005organic}.
Although we employ ${\cal{N}}=4$ supersymmetric Yang-Mills theory with ${\cal{N}}=2$ hypermultiplets in the large-$N_{c}$ limit as our microscopic theory, the difference of the microscopic details between our theory and those in the experimental setups may not affect the description of the critical phenomena, if the idea of universality holds there.
The first-order phase transition observed in \cite{Sawano2005organic} is qualitatively similar to our first-order phase transitions shown in figure \ref{fig:JE} in the small $\rho$ region.
Then, we expect that the second-order phase transition we have seen in the present work may be experimentally observed in the organic conductor in the region of higher carrier density, corresponding to the regime with a large $\rho$ in our system. The large-$N_c$ condition may be relaxed in the experimental setups and the critical exponents can differ from the mean-field values if it is the case.
It is interesting to experimentally search the electric-field-driven second-order phase transitions and determine the values of the critical exponents to study the universality in the nonequilibrium phase transitions.

\section*{Acknowledgements}
The authors thank the fruitful discussion with S. Uemura. The work of S. N. is supported in part by JSPS KAKENHI Grants No. JP19K03659, No. JP19H05821, and the Chuo University Personal Research Grant. 
M.~M. is supported by National Natural Science Foundation of China with Grant No.~12047538.
The authors also thank RIKEN iTHEMS NEW working group for fruitful discussions.

\appendix
\section{Relationship between \texorpdfstring{$J$}{TEXT} and \texorpdfstring{$\rho$}{TEXT}}
\label{AppendixA}
The Lagrangian density is given by
\begin{align}
    {\cal{L}}_{D7} &= -\sqrt{-\det \left(g_{ab} + F_{ab} \right)} \nonumber\\
    &= -\sqrt{-g} \left( 1+ g^{xx}g^{uu} h'^{2} +g^{tt}g^{xx} E^{2} +g^{tt} g^{uu} A_{t}'^{2} \right)^{1/2},
\end{align}
where $g=\det g_{ab}$.
As defined in the main text, we obtain the two conserved quantities 
\begin{align}
    \rho &= \frac{\partial {\cal{L}}_{D7}}{\partial A_{t}'} = \frac{-\sqrt{-g}g^{tt}g^{uu}A_{t}'}{\sqrt{1+ g^{xx}g^{uu} h'^{2} +g^{tt}g^{xx} E^{2} +g^{tt} g^{uu} A_{t}'^{2} }}, \\
    J &= \frac{\partial {\cal{L}}_{D7}}{\partial h'} = \frac{-\sqrt{-g}g^{xx}g^{uu}h'}{\sqrt{1+ g^{xx}g^{uu} h'^{2} +g^{tt}g^{xx} E^{2} +g^{tt} g^{uu} A_{t}'^{2} }}, \label{eq:currentAP}
\end{align}
corresponding to the charge density and current density, respectively.
Combining them, we obtain the following relation
\begin{equation}
    J g^{tt} A_{t}' - \rho g^{xx} h' =0,
\end{equation}
which indicates that $A_{t}'$ and $h'$ are not independent of each other.
By eliminating $A_{t}'$ with this relation in (\ref{eq:currentAP}), we find
\begin{equation}
    J = \frac{-\sqrt{-g}g^{xx}g^{uu}h'}{\sqrt{1 +g^{tt}g^{xx} E^{2} + g^{xx}g^{uu} h'^{2} \left( 1+ \frac{g^{xx}}{g^{tt}}  \frac{\rho^{2}}{J^{2}} \right) }}.
\end{equation}
If we evaluate this at the effective horizon, namely $1+g^{tt}g^{xx}E^{2}=0$,
the gauge field $h'$ no longer appears in the expression and we obtain
\begin{equation}
    J = \left. -\sqrt{-g} \sqrt{\frac{g^{xx}g^{uu}}{1+ \frac{g^{xx}}{g^{tt}}  \frac{\rho^{2}}{J^{2}}}} \right|_{u=u_{*}},
\end{equation}
corresponding to the relationship between the current density and the charge density, i.e. eq.~(\ref{eq:current}) in the main text.

\bibliography{main}

\begin{thebibliography}{24}%
\makeatletter
\providecommand \@ifxundefined [1]{%
 \@ifx{#1\undefined}
}%
\providecommand \@ifnum [1]{%
 \ifnum #1\expandafter \@firstoftwo
 \else \expandafter \@secondoftwo
 \fi
}%
\providecommand \@ifx [1]{%
 \ifx #1\expandafter \@firstoftwo
 \else \expandafter \@secondoftwo
 \fi
}%
\providecommand \natexlab [1]{#1}%
\providecommand \enquote  [1]{``#1''}%
\providecommand \bibnamefont  [1]{#1}%
\providecommand \bibfnamefont [1]{#1}%
\providecommand \citenamefont [1]{#1}%
\providecommand \href@noop [0]{\@secondoftwo}%
\providecommand \href [0]{\begingroup \@sanitize@url \@href}%
\providecommand \@href[1]{\@@startlink{#1}\@@href}%
\providecommand \@@href[1]{\endgroup#1\@@endlink}%
\providecommand \@sanitize@url [0]{\catcode `\\12\catcode `\$12\catcode
  `\&12\catcode `\#12\catcode `\^12\catcode `\_12\catcode `\%12\relax}%
\providecommand \@@startlink[1]{}%
\providecommand \@@endlink[0]{}%
\providecommand \url  [0]{\begingroup\@sanitize@url \@url }%
\providecommand \@url [1]{\endgroup\@href {#1}{\urlprefix }}%
\providecommand \urlprefix  [0]{URL }%
\providecommand \Eprint [0]{\href }%
\providecommand \doibase [0]{https://doi.org/}%
\providecommand \selectlanguage [0]{\@gobble}%
\providecommand \bibinfo  [0]{\@secondoftwo}%
\providecommand \bibfield  [0]{\@secondoftwo}%
\providecommand \translation [1]{[#1]}%
\providecommand \BibitemOpen [0]{}%
\providecommand \bibitemStop [0]{}%
\providecommand \bibitemNoStop [0]{.\EOS\space}%
\providecommand \EOS [0]{\spacefactor3000\relax}%
\providecommand \BibitemShut  [1]{\csname bibitem#1\endcsname}%
\let\auto@bib@innerbib\@empty
\bibitem [{\citenamefont {Oono}\ and\ \citenamefont
  {Paniconi}(1998)}]{Oono1998steady}%
  \BibitemOpen
  \bibfield  {author} {\bibinfo {author} {\bibfnamefont {Y.}~\bibnamefont
  {Oono}}\ and\ \bibinfo {author} {\bibfnamefont {M.}~\bibnamefont
  {Paniconi}},\ }\bibfield  {title} {\bibinfo {title} {Steady state
  thermodynamics},\ }\href@noop {} {\bibfield  {journal} {\bibinfo  {journal}
  {Progress of Theoretical Physics Supplement}\ }\textbf {\bibinfo {volume}
  {130}},\ \bibinfo {pages} {29} (\bibinfo {year} {1998})}\BibitemShut
  {NoStop}%
\bibitem [{\citenamefont {Sasa}\ and\ \citenamefont
  {Tasaki}(2006)}]{Sasa2006steady}%
  \BibitemOpen
  \bibfield  {author} {\bibinfo {author} {\bibfnamefont {S.-i.}\ \bibnamefont
  {Sasa}}\ and\ \bibinfo {author} {\bibfnamefont {H.}~\bibnamefont {Tasaki}},\
  }\bibfield  {title} {\bibinfo {title} {Steady state thermodynamics},\
  }\href@noop {} {\bibfield  {journal} {\bibinfo  {journal} {Journal of
  statistical physics}\ }\textbf {\bibinfo {volume} {125}},\ \bibinfo {pages}
  {125} (\bibinfo {year} {2006})}\BibitemShut {NoStop}%
\bibitem [{\citenamefont {Nakagawa}\ and\ \citenamefont
  {Sasa}(2017)}]{Sasa-Nakagawa}%
  \BibitemOpen
  \bibfield  {author} {\bibinfo {author} {\bibfnamefont {N.}~\bibnamefont
  {Nakagawa}}\ and\ \bibinfo {author} {\bibfnamefont {S.-i.}\ \bibnamefont
  {Sasa}},\ }\bibfield  {title} {\bibinfo {title} {Liquid-gas transitions in
  steady heat conduction},\ }\href
  {https://doi.org/10.1103/PhysRevLett.119.260602} {\bibfield  {journal}
  {\bibinfo  {journal} {Phys. Rev. Lett.}\ }\textbf {\bibinfo {volume} {119}},\
  \bibinfo {pages} {260602} (\bibinfo {year} {2017})}\BibitemShut {NoStop}%
\bibitem [{\citenamefont {Nakagawa}\ and\ \citenamefont
  {Sasa}(2019)}]{Nakagawa2019global}%
  \BibitemOpen
  \bibfield  {author} {\bibinfo {author} {\bibfnamefont {N.}~\bibnamefont
  {Nakagawa}}\ and\ \bibinfo {author} {\bibfnamefont {S.-i.}\ \bibnamefont
  {Sasa}},\ }\bibfield  {title} {\bibinfo {title} {Global thermodynamics for
  heat conduction systems},\ }\href@noop {} {\bibfield  {journal} {\bibinfo
  {journal} {Journal of Statistical Physics}\ }\textbf {\bibinfo {volume}
  {177}},\ \bibinfo {pages} {825} (\bibinfo {year} {2019})}\BibitemShut
  {NoStop}%
\bibitem [{\citenamefont {Sawano}\ \emph {et~al.}(2005)\citenamefont {Sawano},
  \citenamefont {Terasaki}, \citenamefont {Mori}, \citenamefont {Mori},
  \citenamefont {Watanabe}, \citenamefont {Ikeda}, \citenamefont {Nogami},\
  and\ \citenamefont {Noda}}]{Sawano2005organic}%
  \BibitemOpen
  \bibfield  {author} {\bibinfo {author} {\bibfnamefont {F.}~\bibnamefont
  {Sawano}}, \bibinfo {author} {\bibfnamefont {I.}~\bibnamefont {Terasaki}},
  \bibinfo {author} {\bibfnamefont {H.}~\bibnamefont {Mori}}, \bibinfo {author}
  {\bibfnamefont {T.}~\bibnamefont {Mori}}, \bibinfo {author} {\bibfnamefont
  {M.}~\bibnamefont {Watanabe}}, \bibinfo {author} {\bibfnamefont
  {N.}~\bibnamefont {Ikeda}}, \bibinfo {author} {\bibfnamefont
  {Y.}~\bibnamefont {Nogami}},\ and\ \bibinfo {author} {\bibfnamefont
  {Y.}~\bibnamefont {Noda}},\ }\bibfield  {title} {\bibinfo {title} {An organic
  thyristor},\ }\href@noop {} {\bibfield  {journal} {\bibinfo  {journal}
  {Nature}\ }\textbf {\bibinfo {volume} {437}},\ \bibinfo {pages} {522}
  (\bibinfo {year} {2005})}\BibitemShut {NoStop}%
\bibitem [{\citenamefont {Karch}\ and\ \citenamefont
  {Katz}(2002)}]{Karch:2002sh}%
  \BibitemOpen
  \bibfield  {author} {\bibinfo {author} {\bibfnamefont {A.}~\bibnamefont
  {Karch}}\ and\ \bibinfo {author} {\bibfnamefont {E.}~\bibnamefont {Katz}},\
  }\bibfield  {title} {\bibinfo {title} {{Adding flavor to AdS / CFT}},\ }\href
  {https://doi.org/10.1088/1126-6708/2002/06/043} {\bibfield  {journal}
  {\bibinfo  {journal} {JHEP}\ }\textbf {\bibinfo {volume} {06}},\ \bibinfo
  {pages} {043}},\ \Eprint {https://arxiv.org/abs/hep-th/0205236}
  {arXiv:hep-th/0205236} \BibitemShut {NoStop}%
\bibitem [{\citenamefont {Maldacena}(1998)}]{Maldacena:1998}%
  \BibitemOpen
  \bibfield  {author} {\bibinfo {author} {\bibfnamefont {J.~M.}\ \bibnamefont
  {Maldacena}},\ }\bibfield  {title} {\bibinfo {title} {The {large N} limit of
  superconformal field theories and supergravity},\ }\href@noop {} {\bibfield
  {journal} {\bibinfo  {journal} {Adv. Theor. Math. Phys.}\ }\textbf {\bibinfo
  {volume} {2}},\ \bibinfo {pages} {231} (\bibinfo {year} {1998})},\ \Eprint
  {https://arxiv.org/abs/hep-th/9711200} {arXiv:hep-th/9711200} \BibitemShut
  {NoStop}%
\bibitem [{\citenamefont {Gubser}\ \emph {et~al.}(1998)\citenamefont {Gubser},
  \citenamefont {Klebanov},\ and\ \citenamefont {Polyakov}}]{Gubser:1998}%
  \BibitemOpen
  \bibfield  {author} {\bibinfo {author} {\bibfnamefont {S.~S.}\ \bibnamefont
  {Gubser}}, \bibinfo {author} {\bibfnamefont {I.~R.}\ \bibnamefont
  {Klebanov}},\ and\ \bibinfo {author} {\bibfnamefont {A.~M.}\ \bibnamefont
  {Polyakov}},\ }\bibfield  {title} {\bibinfo {title} {Gauge theory correlators
  from non-critical string theory},\ }\href@noop {} {\bibfield  {journal}
  {\bibinfo  {journal} {Phys. Lett. B}\ }\textbf {\bibinfo {volume} {428}}
  (\bibinfo {year} {1998})},\ \Eprint {https://arxiv.org/abs/hep-th/9802109}
  {arXiv:hep-th/9802109} \BibitemShut {NoStop}%
\bibitem [{\citenamefont {Witten}(1998)}]{Witten:1998}%
  \BibitemOpen
  \bibfield  {author} {\bibinfo {author} {\bibfnamefont {E.}~\bibnamefont
  {Witten}},\ }\bibfield  {title} {\bibinfo {title} {Anti-de sitter space and
  holography},\ }\href@noop {} {\bibfield  {journal} {\bibinfo  {journal} {Adv.
  Theor. Math. Phys.}\ }\textbf {\bibinfo {volume} {2}} (\bibinfo {year}
  {1998})},\ \Eprint {https://arxiv.org/abs/hep-th/9802150}
  {arXiv:hep-th/9802150} \BibitemShut {NoStop}%
\bibitem [{\citenamefont {Hubeny}\ and\ \citenamefont
  {Rangamani}(2010)}]{Hubeny:2010ry}%
  \BibitemOpen
  \bibfield  {author} {\bibinfo {author} {\bibfnamefont {V.~E.}\ \bibnamefont
  {Hubeny}}\ and\ \bibinfo {author} {\bibfnamefont {M.}~\bibnamefont
  {Rangamani}},\ }\bibfield  {title} {\bibinfo {title} {{A Holographic view on
  physics out of equilibrium}},\ }\href {https://doi.org/10.1155/2010/297916}
  {\bibfield  {journal} {\bibinfo  {journal} {Adv. High Energy Phys.}\ }\textbf
  {\bibinfo {volume} {2010}},\ \bibinfo {pages} {297916} (\bibinfo {year}
  {2010})},\ \Eprint {https://arxiv.org/abs/1006.3675} {arXiv:1006.3675
  [hep-th]} \BibitemShut {NoStop}%
\bibitem [{\citenamefont {Kundu}(2019)}]{Kundu:2019ull}%
  \BibitemOpen
  \bibfield  {author} {\bibinfo {author} {\bibfnamefont {A.}~\bibnamefont
  {Kundu}},\ }\bibfield  {title} {\bibinfo {title} {{Steady States, Thermal
  Physics, and Holography}},\ }\href {https://doi.org/10.1155/2019/2635917}
  {\bibfield  {journal} {\bibinfo  {journal} {Adv. High Energy Phys.}\ }\textbf
  {\bibinfo {volume} {2019}},\ \bibinfo {pages} {2635917} (\bibinfo {year}
  {2019})}\BibitemShut {NoStop}%
\bibitem [{\citenamefont {Karch}\ and\ \citenamefont
  {O'Bannon}(2007)}]{Karch:2007pd}%
  \BibitemOpen
  \bibfield  {author} {\bibinfo {author} {\bibfnamefont {A.}~\bibnamefont
  {Karch}}\ and\ \bibinfo {author} {\bibfnamefont {A.}~\bibnamefont
  {O'Bannon}},\ }\bibfield  {title} {\bibinfo {title} {{Metallic AdS/CFT}},\
  }\href {https://doi.org/10.1088/1126-6708/2007/09/024} {\bibfield  {journal}
  {\bibinfo  {journal} {JHEP}\ }\textbf {\bibinfo {volume} {09}},\ \bibinfo
  {pages} {024}},\ \Eprint {https://arxiv.org/abs/0705.3870} {arXiv:0705.3870
  [hep-th]} \BibitemShut {NoStop}%
\bibitem [{\citenamefont {Nakamura}(2010)}]{Nakamura:2010zd}%
  \BibitemOpen
  \bibfield  {author} {\bibinfo {author} {\bibfnamefont {S.}~\bibnamefont
  {Nakamura}},\ }\bibfield  {title} {\bibinfo {title} {{Negative Differential
  Resistivity from Holography}},\ }\href {https://doi.org/10.1143/PTP.124.1105}
  {\bibfield  {journal} {\bibinfo  {journal} {Prog. Theor. Phys.}\ }\textbf
  {\bibinfo {volume} {124}},\ \bibinfo {pages} {1105} (\bibinfo {year}
  {2010})},\ \Eprint {https://arxiv.org/abs/1006.4105} {arXiv:1006.4105
  [hep-th]} \BibitemShut {NoStop}%
\bibitem [{\citenamefont {Nakamura}(2012)}]{Nakamura:2012ae}%
  \BibitemOpen
  \bibfield  {author} {\bibinfo {author} {\bibfnamefont {S.}~\bibnamefont
  {Nakamura}},\ }\bibfield  {title} {\bibinfo {title} {{Nonequilibrium Phase
  Transitions and Nonequilibrium Critical Point from AdS/CFT}},\ }\href
  {https://doi.org/10.1103/PhysRevLett.109.120602} {\bibfield  {journal}
  {\bibinfo  {journal} {Phys. Rev. Lett.}\ }\textbf {\bibinfo {volume} {109}},\
  \bibinfo {pages} {120602} (\bibinfo {year} {2012})},\ \Eprint
  {https://arxiv.org/abs/1204.1971} {arXiv:1204.1971 [hep-th]} \BibitemShut
  {NoStop}%
\bibitem [{\citenamefont {Matsumoto}\ and\ \citenamefont
  {Nakamura}(2018)}]{Matsumoto:2018ukk}%
  \BibitemOpen
  \bibfield  {author} {\bibinfo {author} {\bibfnamefont {M.}~\bibnamefont
  {Matsumoto}}\ and\ \bibinfo {author} {\bibfnamefont {S.}~\bibnamefont
  {Nakamura}},\ }\bibfield  {title} {\bibinfo {title} {{Critical Exponents of
  Nonequilibrium Phase Transitions in AdS/CFT Correspondence}},\ }\href
  {https://doi.org/10.1103/PhysRevD.98.106027} {\bibfield  {journal} {\bibinfo
  {journal} {Phys. Rev. D}\ }\textbf {\bibinfo {volume} {98}},\ \bibinfo
  {pages} {106027} (\bibinfo {year} {2018})},\ \Eprint
  {https://arxiv.org/abs/1804.10124} {arXiv:1804.10124 [hep-th]} \BibitemShut
  {NoStop}%
\bibitem [{\citenamefont {Hashimoto}\ \emph {et~al.}(2014)\citenamefont
  {Hashimoto}, \citenamefont {Kinoshita}, \citenamefont {Murata},\ and\
  \citenamefont {Oka}}]{Hashimoto:2014yza}%
  \BibitemOpen
  \bibfield  {author} {\bibinfo {author} {\bibfnamefont {K.}~\bibnamefont
  {Hashimoto}}, \bibinfo {author} {\bibfnamefont {S.}~\bibnamefont
  {Kinoshita}}, \bibinfo {author} {\bibfnamefont {K.}~\bibnamefont {Murata}},\
  and\ \bibinfo {author} {\bibfnamefont {T.}~\bibnamefont {Oka}},\ }\bibfield
  {title} {\bibinfo {title} {{Electric Field Quench in AdS/CFT}},\ }\href
  {https://doi.org/10.1007/JHEP09(2014)126} {\bibfield  {journal} {\bibinfo
  {journal} {JHEP}\ }\textbf {\bibinfo {volume} {09}},\ \bibinfo {pages}
  {126}},\ \Eprint {https://arxiv.org/abs/1407.0798} {arXiv:1407.0798 [hep-th]}
  \BibitemShut {NoStop}%
\bibitem [{\citenamefont {Ishigaki}\ \emph {et~al.}(2022)\citenamefont
  {Ishigaki}, \citenamefont {Kinoshita},\ and\ \citenamefont
  {Matsumoto}}]{Ishigaki:2021vyv}%
  \BibitemOpen
  \bibfield  {author} {\bibinfo {author} {\bibfnamefont {S.}~\bibnamefont
  {Ishigaki}}, \bibinfo {author} {\bibfnamefont {S.}~\bibnamefont
  {Kinoshita}},\ and\ \bibinfo {author} {\bibfnamefont {M.}~\bibnamefont
  {Matsumoto}},\ }\bibfield  {title} {\bibinfo {title} {{Dynamical stability
  and filamentary instability in holographic conductors}},\ }\href
  {https://doi.org/10.1007/JHEP04(2022)173} {\bibfield  {journal} {\bibinfo
  {journal} {JHEP}\ }\textbf {\bibinfo {volume} {04}},\ \bibinfo {pages}
  {173}},\ \Eprint {https://arxiv.org/abs/2112.11677} {arXiv:2112.11677
  [hep-th]} \BibitemShut {NoStop}%
\bibitem [{\citenamefont {Kim}\ \emph {et~al.}(2011)\citenamefont {Kim},
  \citenamefont {Shock},\ and\ \citenamefont {Tarrio}}]{Kim:2011qh}%
  \BibitemOpen
  \bibfield  {author} {\bibinfo {author} {\bibfnamefont {K.-Y.}\ \bibnamefont
  {Kim}}, \bibinfo {author} {\bibfnamefont {J.~P.}\ \bibnamefont {Shock}},\
  and\ \bibinfo {author} {\bibfnamefont {J.}~\bibnamefont {Tarrio}},\
  }\bibfield  {title} {\bibinfo {title} {{The open string membrane paradigm
  with external electromagnetic fields}},\ }\href
  {https://doi.org/10.1007/JHEP06(2011)017} {\bibfield  {journal} {\bibinfo
  {journal} {JHEP}\ }\textbf {\bibinfo {volume} {06}},\ \bibinfo {pages}
  {017}},\ \Eprint {https://arxiv.org/abs/1103.4581} {arXiv:1103.4581 [hep-th]}
  \BibitemShut {NoStop}%
\bibitem [{\citenamefont {Seiberg}\ and\ \citenamefont
  {Witten}(1999)}]{Seiberg:1999vs}%
  \BibitemOpen
  \bibfield  {author} {\bibinfo {author} {\bibfnamefont {N.}~\bibnamefont
  {Seiberg}}\ and\ \bibinfo {author} {\bibfnamefont {E.}~\bibnamefont
  {Witten}},\ }\bibfield  {title} {\bibinfo {title} {{String theory and
  noncommutative geometry}},\ }\href
  {https://doi.org/10.1088/1126-6708/1999/09/032} {\bibfield  {journal}
  {\bibinfo  {journal} {JHEP}\ }\textbf {\bibinfo {volume} {09}},\ \bibinfo
  {pages} {032}},\ \Eprint {https://arxiv.org/abs/hep-th/9908142}
  {arXiv:hep-th/9908142} \BibitemShut {NoStop}%
\bibitem [{\citenamefont {Ishigaki}\ and\ \citenamefont
  {Nakamura}(2020)}]{Ishigaki:2020coe}%
  \BibitemOpen
  \bibfield  {author} {\bibinfo {author} {\bibfnamefont {S.}~\bibnamefont
  {Ishigaki}}\ and\ \bibinfo {author} {\bibfnamefont {S.}~\bibnamefont
  {Nakamura}},\ }\bibfield  {title} {\bibinfo {title} {{Mechanism for negative
  differential conductivity in holographic conductors}},\ }\href
  {https://doi.org/10.1007/JHEP12(2020)124} {\bibfield  {journal} {\bibinfo
  {journal} {JHEP}\ }\textbf {\bibinfo {volume} {12}},\ \bibinfo {pages}
  {124}},\ \Eprint {https://arxiv.org/abs/2008.00904} {arXiv:2008.00904
  [hep-th]} \BibitemShut {NoStop}%
\bibitem [{\citenamefont {Imaizumi}\ \emph {et~al.}(2020)\citenamefont
  {Imaizumi}, \citenamefont {Matsumoto},\ and\ \citenamefont
  {Nakamura}}]{Imaizumi:2019byu}%
  \BibitemOpen
  \bibfield  {author} {\bibinfo {author} {\bibfnamefont {T.}~\bibnamefont
  {Imaizumi}}, \bibinfo {author} {\bibfnamefont {M.}~\bibnamefont
  {Matsumoto}},\ and\ \bibinfo {author} {\bibfnamefont {S.}~\bibnamefont
  {Nakamura}},\ }\bibfield  {title} {\bibinfo {title} {{Current Driven
  Tricritical Point in Large- $N_c$ Gauge Theory}},\ }\href
  {https://doi.org/10.1103/PhysRevLett.124.191603} {\bibfield  {journal}
  {\bibinfo  {journal} {Phys. Rev. Lett.}\ }\textbf {\bibinfo {volume} {124}},\
  \bibinfo {pages} {191603} (\bibinfo {year} {2020})},\ \Eprint
  {https://arxiv.org/abs/1911.06262} {arXiv:1911.06262 [hep-th]} \BibitemShut
  {NoStop}%
\bibitem [{\citenamefont {Matsumoto}(2022)}]{Matsumoto:2022psr}%
  \BibitemOpen
  \bibfield  {author} {\bibinfo {author} {\bibfnamefont {M.}~\bibnamefont
  {Matsumoto}},\ }\bibfield  {title} {\bibinfo {title} {{Tricritical phenomena
  in holographic chiral phase transitions}},\ }\href
  {https://doi.org/10.1007/JHEP11(2022)107} {\bibfield  {journal} {\bibinfo
  {journal} {JHEP}\ }\textbf {\bibinfo {volume} {11}},\ \bibinfo {pages}
  {107}},\ \Eprint {https://arxiv.org/abs/2208.02605} {arXiv:2208.02605
  [hep-th]} \BibitemShut {NoStop}%
\bibitem [{\citenamefont {Karch}\ \emph {et~al.}(2006)\citenamefont {Karch},
  \citenamefont {O'Bannon},\ and\ \citenamefont {Skenderis}}]{Karch:2005ms}%
  \BibitemOpen
  \bibfield  {author} {\bibinfo {author} {\bibfnamefont {A.}~\bibnamefont
  {Karch}}, \bibinfo {author} {\bibfnamefont {A.}~\bibnamefont {O'Bannon}},\
  and\ \bibinfo {author} {\bibfnamefont {K.}~\bibnamefont {Skenderis}},\
  }\bibfield  {title} {\bibinfo {title} {{Holographic renormalization of probe
  D-branes in AdS/CFT}},\ }\href
  {https://doi.org/10.1088/1126-6708/2006/04/015} {\bibfield  {journal}
  {\bibinfo  {journal} {JHEP}\ }\textbf {\bibinfo {volume} {04}},\ \bibinfo
  {pages} {015}},\ \Eprint {https://arxiv.org/abs/hep-th/0512125}
  {arXiv:hep-th/0512125} \BibitemShut {NoStop}%
\bibitem [{\citenamefont {Matsumoto}\ and\ \citenamefont
  {Nakamura}(2022)}]{Matsumoto:2022nqu}%
  \BibitemOpen
  \bibfield  {author} {\bibinfo {author} {\bibfnamefont {M.}~\bibnamefont
  {Matsumoto}}\ and\ \bibinfo {author} {\bibfnamefont {S.}~\bibnamefont
  {Nakamura}},\ }\bibfield  {title} {\bibinfo {title} {{Current-induced inverse
  symmetry breaking and asymmetric critical phenomena at current-driven
  tricritical point}},\ }\href {https://doi.org/10.1103/PhysRevD.106.026006}
  {\bibfield  {journal} {\bibinfo  {journal} {Phys. Rev. D}\ }\textbf {\bibinfo
  {volume} {106}},\ \bibinfo {pages} {026006} (\bibinfo {year} {2022})},\
  \Eprint {https://arxiv.org/abs/2201.06894} {arXiv:2201.06894 [hep-th]}
  \BibitemShut {NoStop}%
\end{thebibliography}%
\end{document}